\documentclass[12pt,preprint]{aastex}  

\usepackage{epsfig}
\usepackage{natbib}

\shorttitle{MHD Jet}

\shortauthors{Li et al.}

\begin{document}

\title{Modeling the Large Scale Structures of Astrophysical Jets in
  the Magnetically Dominated Limit}

\author{Hui Li\altaffilmark{1}, Giovanni Lapenta\altaffilmark{2}, 
John M. Finn\altaffilmark{2}, Shengtai Li\altaffilmark{3}, and 
Stirling A. Colgate\altaffilmark{4}}
\altaffiltext{1}{
Applied Physics Division, MS B227, Los Alamos National
Laboratory, NM 87545; hli@lanl.gov}
\altaffiltext{2}{
Theoretical Division, MS K717, Los Alamos National
Laboratory, NM 87545; lapenta@lanl.gov; finn@lanl.gov}
\altaffiltext{3}{Theoretical Division, MS B284, Los Alamos National
Laboratory, NM 87545; sli@lanl.gov}
\altaffiltext{4}{Theoretical Division, MS B227, Los Alamos National
Laboratory, NM 87545; colgate@lanl.gov}

\begin{abstract}
We suggest a new approach that could be used for modeling both the
large scale behavior of astrophysical jets and the magnetically
dominated explosions in astrophysics.  We describe a method for
modeling the injection of magnetic fields and their subsequent
evolution in a regime where the free energy is magnetically
dominated. The injected magnetic fields, along with their associated
currents, have both poloidal and toroidal components, and they are not
force free.  The dynamic expansion driven by the Lorentz force of the
injected fields is studied using 3-dimensional ideal
magnetohydrodynamic simulations.  The generic behavior of magnetic
field expansion, the interactions with the background medium, and the
dependence on various parameters are investigated.
\end{abstract}



\section{INTRODUCTION}
\label{sec:intro}

It is generally thought that energies (including magnetic fields) in
jets and lobes of radio galaxies observed on large scales (from kpc to
Mpc) are supplied from the central black hole accretion region,
despite our poor understanding on how exactly this has occurred. Near
the black hole, studies have focused on the initial formation of jets,
in terms of jet collimation and acceleration. Here, both hydromagnetic
limit (e.g., Blandford \& Payne 1982; Ouyed \& Pudritz 1997; Ustyugova
et al. 1999; and others) and Poynting flux limit (e.g., Blandford
1976; Lovelace 1976; Lynden-Bell 1996; Ustyugova et al. 2000; Li et
al. 2001; Lovelace et al. 2002; and others) have been investigated,
with various degree of success [see Ferrari (1998) for a relatively
recent review].

On much larger scales, jet propagation, morphologies (such as lobe
formation, hot-spots, filaments, etc.) and associated
instabilities have been investigated extensively. In particular, much
progress has been made in the so-called kinetic energy dominated
regime, for both non-relativistic and relativistic flows. The
pioneering work by Norman et al. (1988) set the stage for nearly two
decades of studies on how highly supersonic, light (jet material
density less than the surrounding density), and sometimes magnetized,
collimated flows propagate in various (uniform and/or stratified)
background medium (e.g., Norman et al. 1983; Mart\'i et al. 1994;
Clarke 1996; Bodo et al. 1998; Tregillis et al. 2004; and many
others). Even though magnetic fields have been actively included in
some of these studies, the overall behavior has been predominantly
determined in the hydrodynamic limit (but see a recent paper by
Nakamura \& Meier 2004 in both hydro and Poynting dominated limits).

The relative importance of kinetic energy and magnetic energy in jets
and radio lobes has been a subject for extensive studies and debates.
Better radio observations on large, $>$ tens of kpc, scales (e.g.,
Owen et al. 2000) and recent X-ray observations of radio lobes of
Fanaroff-Riley type II (FR II) sources (e.g., Croston et al. 2005)
have suggested that magnetic energies inside the radio lobes are
significant (see also Kronberg et al. 2001). These observations bring
out the questions whether jets/lobes could be magnetically dominated and
what are the physical processes govern such flows on these large scales.
The studies on this Poynting flux dominated limit in large scales have
been relatively sparse.

The purpose of the present paper is to present an approach to model
the large scale magnetic structure generated by a purely magnetic
energy input. We describe a method of injecting magnetic fields into a
small volume and following the subsequent evolution of the injected
magnetic free energy. This free energy causes magnetic fields to
expand under their own Lorentz force. This expansion is subsequently
affected by interacting with the background environments. We believe
that this approach could in principle be applied to different
astrophysical systems. Here, we concentrate on its applications to
modeling the large scale morphologies of radio jets and lobes.  In \S
\ref{sec:model}, we describe our basic assumptions and approach on how
to model magnetic field injection. Numerical methods used in our
simulations are also discussed. In \S \ref{sec:resul}, we present the
simulation results on the evolution of magnetic fields for different
parameters and setups.  In \S \ref{sec:discu}, we summarize our
results.

\section{Basic Approach and Model Assumptions}
\label{sec:model}

Since it is currently impossible to simulate the whole process from
the jet formation near the black hole accretion disk out to radio
lobes because of the vast length scale separation, we adopt the
following general framework which separates the jet/lobe structure
into different regions: I: the engine region, where the black hole
accretion disk system produces magnetic fields and these fields are
ejected above and below the disk. II: the propagation region, where
energy, mass, and magnetic fields flow through the background medium.
III: the termination region, where these flows are significantly
slowed down or come to a stop. In our approach, for region I, we
emphasize that we are {\it not} modeling the black hole accretion disk
system and how magnetic fields are ejected from such systems.
Instead, we attempt to model the likely {\it outcome} of the
energy/mass supply from the black hole accretion system as having
particular functional forms in non-force-free magnetic fields, along
with certain injection rates. In this sense, we are focusing on the
energy/mass flows in an external medium, and how they would eventually
terminate (i.e.  Regions II and III).  In particular, we study the
limit that the injected free energy is predominantly in magnetic
fields, which is different from the previous approach where the flow
is dominated by the kinetic energy (i.e., the injected kinetic energy
is much larger than the injected magnetic and thermal energy).  In the
following sections, we will discuss the functional forms of injected
magnetic fields and the numerical setups for modeling the nonlinear
evolution of these magnetic fields.

\subsection{Basic Equations, Magnetic Fields, Mass Injections
  and Numerical Codes}
\label{sec:code}

We solve the ideal MHD equations in three-dimensional Cartesian
coordinate system $\{x,y,z\}$:
\begin{eqnarray}
\frac{\partial\rho}{\partial t} + \nabla\cdot(\rho\mbox{\bf v}) &=& 
\dot{\rho}_{\rm inj}
\\
\frac{\partial (\rho {\bf v})}{\partial t} + 
\nabla\cdot \left( \rho {\bf v v} + P_{\rm g} + P_{\rm B}- 
{\bf B B}\right)  &=& 0
\\
\frac{\partial E}{\partial t} + \nabla\cdot\left[\left(E
+P_{\rm g}+P_{\rm B}\right)
{\bf v}-{\bf B}({\bf v}\cdot{\bf B})\right] &=&
\dot{E}_{\rm inj} \label{eq:totE}
\\
\frac{\partial {\bf B}}{\partial t} - \nabla\times( {\bf v}
\times {\bf B}) &=&  {\dot {\bf B}}_{\rm inj}~~,
\end{eqnarray}
where all variables have their usual meaning. We use an ideal gas
equation of state with an adiabatic index $\gamma = 5/3$, which
relates the internal energy density $e$ and the gas pressure $P_{\rm
g}$ as $e = P_{\rm g}/(\gamma-1)$. The magnetic energy density is
expressed as $P_{\rm B}=B^2/2$.  The total energy $E = \rho v^2/2 +
P_{\rm g} + P_{\rm B}$. We have neglected external forces such as
gravity in these simulations. This is not correct for some
astrophysical environments such as jets in galaxy clusters. The main
focus of this paper is to present the salient features of this new
approach, and we will present more detailed astrophysical applications
in future publications.

We have added an injection term ${\dot {\bf B}}_{\rm inj}$ in the
induction equation to incorporate the external injection of magnetic
fields ($\nabla \cdot {\dot {\bf B}}_{\rm inj} = 0$).  The injection
process is modeled as:
\begin{equation}
{\dot {\bf B}}_{\rm inj} = \left\{ \begin{array}{cc}
\gamma_b ~{\bf B}_{\rm inj} & {\rm ~for~~} t \leq t_{\rm inj}\\
0                           & {\rm ~for~~} t > t_{\rm inj}
\end{array}\right.
\end{equation} 
where ${\bf B}_{\rm inj}$ describes the magnetic fields whose exact
form is given in the following section.  The $\gamma_b(t)$ specifies
the injection rate for magnetic fields, and it is usually taken as a
constant over a finite time internal $t \leq t_{\rm inj}$ and as zero
(e.g., the injection is turned off) for $t > t_{\rm inj}$.  More
general forms of injection can be easily included as well.

In addition, we have included mass injection in the central region,
motivated by the possibility that matter can be ejected together with
magnetic fields away from the central source. Note that both the
magnetic field and mass injections violate the ideal MHD condition, in
the sense that the total mass on magnetic field lines are allowed to
change.  Furthermore, adding mass unto field lines helps to alleviate
the numerical problems of having extreme low density regions after
magnetic fields have undergone large expansions.  Without the detailed
knowledge on how and what amount of mass could get unto the field
lines, we adopt the following simple formula:
\begin{equation}
\label{eq:rhoinj}
\dot{\rho}_{\rm inj} = \gamma_\rho \rho_0 \exp[-(r^2 + z^2)/r_\rho^2]~,
\end{equation} 
where $r_\rho$ and $\gamma_\rho$ are the characteristic radius and
rate for mass injection. The injected mass is further assumed to have
the same velocity and temperature as that of the gas in the injection
region.

Furthermore, there is an energy injection along with the magnetic
field and mass injections. The energy injection rate from the magnetic
fields injection can be expressed as
\begin{equation}
\dot{E}_{\rm inj} = {\bf B}\cdot {\dot {\bf B}}_{\rm inj} = \gamma_b
 {\bf B}_{\rm inj}\cdot {\bf B}~,
\end{equation} 
while in general both ${\bf B}$ and ${\bf B}_{\rm inj}$ have temporal
and spatial dependences.  Note that there is a small amount of thermal
and kinetic energy injected, accompanied with the mass injection. But
this energy is negligibly small compared with the injected magnetic
energy.

All simulations are performed using a new ideal MHD package developed
at Los Alamos (Li \& Li 2003). This package uses high-order
Godunov-type finite volume numerical methods. These methods
conservatively update the zone-averaged fluid and magnetic field
quantities based on estimated advective fluxes of mass, momentum,
energy and magnetic field at zone interfaces. Due to the
discontinuities (contacts and shocks) in the solutions, Roe's Riemann
solver with an entropy-fix (Powell et al. 1994) is used to calculate
these fluxes. It usually solves the total energy equation as the ideal
MHD equations are written into the conservative form. To maintain
pressure positivity in regions where the pressure can become very
small or negative, we also solve two auxiliary equations: the internal
energy equation and the modified entropy equation,
\begin{eqnarray}
\frac{\partial e}{\partial t} + \nabla\cdot(e\mbox{\bf v}) &=& 
- P_{\rm g}\nabla\cdot\mbox{\bf v} \label{eq:inte}
\\
\frac{\partial S}{\partial t} + \nabla\cdot(S\mbox{\bf v}) &=& 0 
\label{eq:modS}
\end{eqnarray}
where $S = P_{\rm g}/\rho^{(\gamma-1)}$ is defined as the modified
entropy. The exact usage of these two equations is described in Ryu et
al. (1993) and Balsara \& Spicer (1999a).

The divergence free condition of the magnetic field $\nabla\cdot {\bf
  B} = 0$ is ensured by a constrained transport (CT) scheme flux-CT
(Balsara \& Spicer 1999b). The CT scheme can yield magnetic fields that
are different from the updated values based on the Godunov method. The
difference may be large enough to lead to negative pressure. To
minimize the change in energy conservation, we update the total energy
with the newly-calculated magnetic fields only when the pressure
becomes negative.  The whole package is parallelized via
message-passing interface (MPI).  A typical run made on parallel Linux
clusters with 64 processors takes 3 hrs.

\subsection{Injected Magnetic Field Configuration}

We determine the injected magnetic fields using three key quantities:
the length scale of the injection region (which we designate as
$r=1$), the amount of poloidal flux ($\Psi_p$), and the poloidal
current ($I_z$). For simplicity, we also assume that the injected 
magnetic fields ${\bf  B}_{\rm inj}$ are axisymmetric.  
In cylindrical coordinates
$\{r,z,\phi\}$, we specify the injected poloidal flux function $\Psi$
as, (where $\Psi=r A_\phi$ and $A_\phi$ is the $\phi$ component of
vector potential),
\begin{equation}
\label{eq:psi}
\Psi(r,z) = r^2 \exp(-r^2 - z^2)~,
\end{equation}
up to a normalization coefficient $B_0$.
Note that one can easily introduce another scale variable in the
$z-$direction which separates the scaling between radial and vertical
directions. Here, we have taken them to be same for simplicity.  In
the global distributions, we envision that the poloidal flux and
current will point along one direction in one region (say, for $r \leq
1$) and ``return'' in the opposite direction in another region (say,
for $r> 1$), so the total net poloidal flux and current are
zero. 

The above-mentioned key ingredients are motivated by the general
framework of treating the black hole accretion disk system as a dynamo
``machine'', from which we use Eq.(\ref{eq:psi}) as a simple
representation of the axisymmetric part of the poloidal flux that
might come from the dynamo process. The continuous shearing of the
poloidal flux lines by the disk's differential rotation generates the
toroidal magnetic fields which correspond to a large scale poloidal
current. Although the system scale we are modeling here is much larger
than the actual size of the black hole accretion disk system, we
envision that there still exists such global poloidal flux and current
distributions on these scales.

The poloidal fields, up to a normalization coefficient $B_0$, are:
\begin{eqnarray}
B_{{\rm inj},r} &=& -\frac{1}{r} \frac{\partial \Psi}{\partial z} = 2 z r 
\exp(-r^2 - z^2)~~,\label{eq:br}\\
B_{{\rm inj},z} &=& \frac{1}{r} \frac{\partial \Psi}{\partial r} = 2 (1-r^2) 
\exp(-r^2 - z^2)~~. \label{eq:bz}
\end{eqnarray}
The poloidal fields have an $O$-point at $(r,z) = (1,0)$ where both
$B_{{\rm inj},r}$ and $B_{{\rm inj},z}$ are zero. In specifying the
toroidal component $B_{{\rm inj},\phi}$, we adopt the approach of
writing 
\begin{equation}
B_{{\rm inj},\phi} = f(\Psi)/r~,
\end{equation}
where $f(\Psi)$ is an arbitrary function of $\Psi$. Physically, this
means that we have made the $\hat{\phi}-$component of the Lorentz
force $({\bf J} \times {\bf B})_\phi$ of the injected magnetic fields
equal to zero. However, this does not prevent rotation being developed
during the evolution by the Lorentz force of the {\em combined} fields
and currents from both the injected fields and the already existing
fields. Specifically, we choose $f(\Psi) = \alpha \Psi$, which means
\begin{equation}
\label{eq:bt}
B_{{\rm inj},\phi} = \alpha \Psi /r = \alpha r \exp(-r^2 - z^2)~~,
\end{equation}
where $\alpha$ is a constant and it has the units of inverse length
scale. Other forms of $f(\Psi)$ are certainly possible, which will
give different pitch profiles for the magnetic fields.

\begin{figure}[ht]
\caption{{\it left}: Shown is the radial component of the ${\bf
    J}\times {\bf B}$ force as a function of radius $r$ in the
    equatorial plane with $\alpha = 1, 3,$ and $5$ for solid,
    dot-dashed and dashed lines, respectively. {\it right}: Shown is
    the vertical component of the ${\bf J}\times {\bf B}$ force as a
    function of radius $r$ at the $z=0.5$ plane with $\alpha = 1,
    3,$ and $5$ for solid, dot-dashed and dashed lines,
    respectively.}
\label{fig:force}
\end{figure}

This simple initial configuration has several properties. Firstly, the
magnetic fields are ``localized'' because the flux and energy drop
exponentially as a function of distance from the center. This is an
important difference from many previous studies where large scale
background magnetic fields are often assumed to be present initially.
In the poloidal plane, this global dipole-like configuration gives
zero net flux and current.  

Secondly, the parameter $\alpha$ roughly specifies the flux ratio
between toroidal and poloidal components. The poloidal flux between $0
\leq r \leq 1$ is
\begin{equation}
\label{eq:pflx}
\Phi_p = 2\pi\Psi(1,0) = 2\pi/e \approx 2.3~~,
\end{equation} 
and the toroidal flux threading through the $\{x,z\}$ plane with
$x\geq 0$ is
\begin{equation}
\Phi_t = \alpha \sqrt{\pi}/2 \approx 0.9 \alpha~~.
\end{equation} 
So, the poloidal and toroidal fluxes are roughly equal when $\alpha
\approx 2.6$.  In astrophysical jets powered by AGNs, it is likely
that magnetic fields are highly wound up by the disk rotation, this
will imply that $\alpha >>1$.  The poloidal current flowing through
the middle plane ($z=0$) with $0 \leq r \leq 1$ is
\begin{equation}
\label{eq:piz}
I_z = 2\pi f(\Psi(1,0)) = 2\pi \alpha/e \approx 2.3 \alpha~~.
\end{equation} 
The total toroidal current is
\begin{equation}
I_\phi = \int\int J_\phi dr dz = 2 \sqrt{\pi}~~. 
\end{equation} 
The ratio of poloidal ($I_z$) to toroidal current ($I_\phi$) is
$\alpha \sqrt{\pi}/e \approx 0.65 \alpha$, i.e., the poloidal current
becomes dominant when $\alpha \gg 1$.

Thirdly, it is important to realize that this initial field is not in
force equilibrium because ${\bf J}\times {\bf B} \ne 0$. The current
density ${\bf J} = \nabla \times {\bf B_{\rm inj}}$ is given as
\begin{eqnarray}
J_r &=& 2\alpha z r ~\exp(-r^2 - z^2)~~,\\
J_z &=& 2\alpha (1 - r^2)~ \exp(-r^2 - z^2)~~,\\
J_\phi &=& 2 r (5 - 2z^2 - 2r^2)~ \exp(-r^2 - z^2)~~.
\end{eqnarray} 
Along the equatorial plane ($z=0$), the radial force density is given
as
\begin{equation}
F_r(z=0) = 2 r (1 - r^2) (10 - \alpha^2 - 4r^2) \exp(-2r^2)~~.
\end{equation} 

It can be seen that at large $r$, $F_r >0$, meaning that the magnetic
fields always expand outward due to the ``hoop'' forces. At small $r$
and small $\alpha$, the force is radially outward as well, whereas for
small $r$ but large $\alpha$, the force is radially inward (i.e.,
``pinching'').  Figure \ref{fig:force} ({\it left}) illustrates these
effects along the equatorial plane for different $\alpha$
values. Along the vertical direction,
\begin{equation}
F_z = 2 z r^2 (\alpha^2 - 10 + 4z^2 + 4r^2) 
\exp(-2r^2 - 2z^2)~~.
\end{equation} 
For $\alpha^2 > 10$, $F_z$ is always positive (negative) for
$z>0 (z<0)$, driving the fields away from the mid-plane. For smaller
$\alpha$, fields at large spherical radii will still expand away from
the origin. Figure \ref{fig:force} ({\it right}) shows the vertical force as
a function of $r$ at $z=0.5$ for different $\alpha$.

For the sake of completeness, we also give the total magnetic energy
of this field:
\begin{equation}
\label{eq:eb}
E_0 = \int (B_{\rm inj}^2/2) dV =  \sqrt{\frac{\pi}{2}}~
\frac{(5+\alpha^2)\pi}{8}
\approx 0.5 (5 + \alpha^2)~~,
\end{equation} 
out of which the first and second terms are from the poloidal and toroidal
magnetic field components, respectively.  

In summary, we have devised a magnetic field structure that is
governed by the poloidal flux $\Psi$ and poloidal current $I_z$. This
magnetic field is localized in space, containing both poloidal and
toroidal fluxes which are interwoven (i.e., finite helicity), and will
dynamically evolve (i.e., not in force-free equilibrium), especially
along the vertical direction in the large $\alpha$ limit. As we will
show later in detail, it is this vertical expansion along the symmetry
axis that forms the basis of modeling the large scale structure of
jets.

\subsection{Units for 3D Simulations}

The initial magnetic field structure is embedded in a plasma
background with finite gas pressure and density, both of which are
taken to be uniform for simplicity in this study. We take $P_{\rm g} =
P_0$, $\rho=\rho_0$, and the sound speed $c_s = \sqrt{\gamma P_{\rm
g}/\rho} \approx 1.3\sqrt{P_0/\rho_0}$. The magnetic field strength
normalization parameter is $B_0$.  Simulations are performed in the
Cartesian geometry with a uniform mesh. The simulation domain is a
cube and the resolution is usually taken as $240\times 240 \times 240$
or higher.

For the following runs, we typically choose $\rho_0 = 1$, $P_0 = 1$,
and $B_0 = 1$. To put these numbers in an astrophysics context:
suppose that the code unit $\rho_0 = 1$ corresponds to a background
density of $3\times 10^{-3}$ particles/cm$^3$, and the background
temperature is $7$ keV. This means that the sound speed in the code
unit $c_s = \sqrt{\gamma} \approx 1.3$ corresponds to $1.1\times 10^8$
cm/s.  The combined density and temperature relation also means that
the magnetic field $B_0=1$ in the code units corresponds to a physical
magnetic field of $\sim 20 \mu$G.  Suppose the code unit $L=1$
corresponds to $15$ kpc, then the simulation duration $t=5$
corresponds to $\sim 9.2\times 10^7$ yrs, and a total energy of $200$
corresponds to $\sim 6.2\times 10^{59}$ ergs. The poloidal flux as
given in Eq.(\ref{eq:pflx}) is $\sim 9.4\times 10^{40}$ Gcm$^2$, and
the poloidal current as given in Eq.(\ref{eq:piz}) is $\sim
1.7\alpha\times 10^{18}$ Ampere.

\section{Results}
\label{sec:resul}

We present the simulation results for two cases: the first has
impulsively ``injected'' magnetic fields (i.e., as initial conditions)
for $\alpha = 20$. The second case is with a continuous injection of
magnetic fields.

\begin{figure}[ht]
\caption{The impulsive injection case with $\alpha = 20$. ({\it left}):
  Shown is the azimuthally averaged pressure in the $\{r,z\}$
  plane. Azimuthally averaged poloidal flux surfaces (white-color) are
  overlaid as well. They are at $t=0, 0.5, 1$, respectively. ({\it
  right}): Same as in the left panel, except that $t = 3, 4, 5$,
  respectively. Note the cartesian computational box is $[-12,12]$ in
  all three directions for the simulation, and axis limits and
  colorbars have been changed between left and right panels.}
\label{fig:alp20}
\end{figure}

\subsection{Impulsive Injection with an initial $\alpha=20$}

We choose $P_0= \rho_0 = B_0 = 1$ and $\alpha=20$ for this run and the
initial ${\bf B}$ is of the form of ${\bf B}_{\rm inj}$. The injection
coefficients $\gamma_b$ and $\gamma_\rho$ are taken to be zero (i.e.,
no continuous injection of ${\bf B}$ or $\rho$).  The computational box
is taken to be $[-12,12]$ in all three directions.  These parameters
give the maximum initial magnetic field $|B| \approx 8.6$ at $(r,z)
\approx (0.7,0)$.

\begin{figure}[ht]
\caption{The impulsive injection case with $\alpha = 20$. Shown is the
overall energy evolution of different components.  The magnetic,
kinetic and changes in thermal energies are shown as solid, dashed,
dot-dashed curves, respectively. The total energy is shown by the thin
solid curve at the top. As the system evolves, the initial magnetic
energy is converted to both the thermal energy and the kinetic energy
of the plasma.}
\label{fig:enealp20}
\end{figure}

The ``free'' energy of the system is initially all magnetic. Since the
initial magnetic field is out of force equilibrium (see Figure
\ref{fig:force}), evolution is expected. Because the evolution remains
quite axisymmetric (see below), in Figure \ref{fig:alp20}, we show the
time sequence of azimuthally averaged pressure distribution, overlaid
with azimuthally averaged poloidal flux function $\Psi$. In Figure
\ref{fig:enealp20}, we show the overall energy evolution for different
components. The initial magnetic energy is given by Eq.(\ref{eq:eb})
with $\alpha=20$. The kinetic energy is $\rho v^2/2$ integrated over
the whole volume, and the change in thermal energy is $[P_{\rm
g}(t)-P_{\rm g}(0)]/(\gamma-1)$ integrated over the whole volume,
where $P_{\rm g}(t)$ and $P_{\rm g}(0)$ are the pressure distributions
at time $t$ and $t=0$.

The whole evolution can be approximately separated into two stages.
The first stage ($0 < t < 1$) is a rapid conversion of $\sim 70\%$ of
the available magnetic energy through the work done by the ${\bf
J}\times {\bf B}$ force on the surrounding plasmas, which pinches all
the flux lines within $r =1$ radially inward and expands flux lines
vertically away from the mid-plane. Since $\alpha$ in this case is
quite high, the ${\bf J}\times {\bf B}$ force is consequently so
strong that it initially drives both a radially converging shock and
an axially expanding shock, as indicated by the central red region and
the red region surrounding the magnetic fields in the left panel of
Figure \ref{fig:alp20}. These shocks heat the plasmas efficiently and
generate large amounts of entropy. The second stage is much slower,
represented by the slow dissipation of magnetic energy and further
gradual heating of the plasma. The magnetic field expansion gradually
slows down. In the meantime, a compressional shock wave, shown as the
outer red region in the right panel of Figure \ref{fig:alp20},
continues with its expansion. Figures \ref{fig:vrtalp20} and
\ref{fig:pretalp20} show the time evolution of the azimuthally
averaged radial profiles of the radial velocity $v_r$ and the plasma
pressure $P_{\rm g}$. A shock front, generated from the initial
impulsive release of the magnetic energy, can indeed be seen.  This
shock heats the plasmas surrounding the magnetic structure as
indicated by the propagating pressure front. The continued expansion,
however, also makes a central, low pressure cavity. Note that the gas
pressure changes from this wave propagation are quite strong ($\sim
20-40\%$).

\begin{figure}[ht]
\caption{The impulsive injection case with $\alpha = 20$. Shown is the
  radial distribution of the azimuthally averaged (cylindrical) radial
  velocity $v_r$ at $t= 0.5, 1.5, ..., 4.5$, respectively. The
  profiles are taken at $z=2.4$. The front generated by the sudden
  release of magnetic energy is moving supersonically. }
\label{fig:vrtalp20}
\end{figure}

\begin{figure}[ht]
\caption{The impulsive injection case with $\alpha = 20$. Shown is the
  radial distribution of the azimuthally averaged pressure at $t= 0.5,
  1.5, ..., 4.5$, respectively. The profiles are taken at $z=2.4$. At
  early time, the ``pinching'' of inner magnetic fields greatly
  increases the pressure near the central axis. This is followed by
  expansion which creates a central low pressure (and density)
  region. The front is moving supersonically.}
\label{fig:pretalp20}
\end{figure}

\begin{figure}[ht]
\caption{The impulsive injection case with $\alpha = 20$. Shown is the
  magnetic fields $|B|$, density and pressure distributions in the
  $\{x,z\}$ plane ($y=0$) at $t=5$. The overall structure remains
  axisymmetric. The magnetic field expansion creates a central low
  density and low pressure region.}
\label{fig:quanalp20}
\end{figure}

\begin{figure}[ht]
\caption{The impulsive injection case with $\alpha = 20$. Shown is one
  magnetic field line at $t=5$. The field line goes up in a tightly
  wound central helix and comes back in a loosely wound helix.}
\label{fig:flalp20}
\end{figure}

\begin{figure}[ht]
\caption{The impulsive injection case with $\alpha = 20$. ({\em
  Upper}): The radial distribution of magnetic field components $B_z$
  and $B_\phi$. ({\em Lower}): The radial distribution of the plasma
  pressure $P_g$, the magnetic pressure $P_B$, and the total pressure
  $P_{\rm tot} = P_g + P_B$. Both panels are taken at $z=3$ and
  $t=5$. }
\label{fig:radcut20}
\end{figure}

Figure \ref{fig:quanalp20} shows the distribution of the magnitude of
magnetic fields $|B|$, density $\rho$ and gas pressure $P_{\rm g}$ at
$t=5$ in the $\{x,z\}$ plane ($y=0$). These distributions suggest that
the final state is still quite axisymmetric. The magnetic fields form
an elongated structure due to both the pinch at the central region and
the vertical expansion along the symmetry axis.  Figure
\ref{fig:flalp20} depicts a 3D field line, showing its going up in a
central tightly wound helix and coming down outside in a loosely wound
helix. Based on the rate of field twists, one might expect that this
helix should be kink unstable. The hollow pressure profile as shown in
Figure \ref{fig:quanalp20}, however, may be a stabilizing influence.

Figure \ref{fig:radcut20} shows the radial distribution of azimuthally
averaged magnetic field components (top) and different pressure
components (bottom) at $z=3$ and $t=5$.  The expansion of the magnetic
fields from an initial sphere with a radius of $\sim 1$ to the
elongated structure have created a very low density cavity since
plasmas are tied to the expanding magnetic fields. The pressure within
this cavity is mostly from the shock heating. The cavity is dominated
by magnetic pressure (the plasma $\beta=P_{\rm g}/P_B$ is $\sim 10\%$
near the axis). A careful force balance analysis shows that the
central region is nearly force free (more detailed results will be
presented elsewhere).  Going to large radii ($r \sim 1-1.5$), however,
the structure is confined by the gas pressure, which prevents further
expansion of the magnetic fields.

Since the density is quite low inside the cavity, there is some
numerical heating, which amounts to an error of $\sim 2\%$ increase in
the total energy (the top thin solid curve in Figure
\ref{fig:enealp20}).  Some of the poloidal flux lines at late times
(the plot at $t=5$ of Figure \ref{fig:alp20}) indicate numerical
magnetic reconnection at the mid-plane.

\subsection{Continuous Injection}

\begin{figure}[ht]
\caption{The continuous injection case with an injection $\alpha =
  15$. Shown is the evolution of azimuthally averaged poloidal flux
  lines during and after the magnetic field injection. Different plots
  are for $t=0, 2.5, 4, 5.5,$ and $7$. The injection is terminated at
  $t=2.5$. Each subplot has ten, evenly spaced contour lines.}
\label{fig:psicontinj}
\end{figure}

\begin{figure}
\caption{The continuous injection case with an injection $\alpha =
  15$. Shown is the evolution of different energy components. The
  thicker solid, dashed and dot-dashed lines are for magnetic, kinetic
  and changes in thermal energies, respectively. The thinner dashed
  and solid curves at the top are for the total injected energy and
  the total energy in the computational box, respectively. }
\label{fig:enecontinj}
\end{figure}

\begin{figure}
\caption{The continuous injection case with an injection $\alpha =
  15$. Shown is the magnetic field, density and pressure distributions
  (from left to right) at $t=7$. The central low density and pressure
  volume is occupied with and dominated by magnetic fields. A
  compressional shock wave generated during the magnetic field
  injection stage has propagated away from the central, magnetized
  region.}
\label{fig:finalcontinj}
\end{figure}

\begin{figure}
\caption{The continuous injection case with an injection $\alpha =
  15$. ({\em Upper}): The radial distribution of azimuthally averaged
  magnetic field components $B_z$ and $B_\phi$. ({\em Lower}): The
  radial distribution of the plasma pressure $P_g$, the magnetic
  pressure $P_B$, and the total pressure $P_{\rm tot} = P_g +
  P_B$. Both panels are taken at $z=3$ and $t=7$.}
\label{fig:radcutcontinj}
\end{figure}

For this case, we adopt the functional form of ${\bf B}_{\rm inj}$ for
both the initial and injected magnetic fields. For the initial fields,
we choose $P_0 = \rho_0 = 1$, $B_0 = 1$, and $\alpha=3$. For the
continuous field injection, we use $B_0 = 0.2$, $\alpha = 15$,
$\gamma_b = 3$, $\gamma_\rho = 0.1$, $r_\rho = 0.5$, and $t_{\rm inj}
= 2.5$. The computational box is $[-12,12]$ in all three
dimensions. This run uses a grid of $320^3$.  To put these numbers in
an astrophysics context, using the conversions given in the beginning
of this section, we have: the simulation's $t_{\rm inj} = 2.5$
corresponds to $\sim 4.6\times 10^7$ yrs, and a total energy of $200$
corresponds to $\sim 6.2\times 10^{59}$ ergs.  The mass injection rate
$0.1$ corresponds to $\sim 1 M_\odot/yr$ and the magnetic field
$B_0=0.2$ corresponds to $\sim 4 \mu$G.  All these parameters are
roughly consistent with the physical conditions in the central region
of galaxy clusters.  However, note that the background medium is not
stratified in the current simulation, so the total thermal energy
within a computational volume of $10^3$ would correspond to $\sim
3\times 10^{60}$ ergs. In other words, the magnetic fields are
expanding into a volume with a high thermal pressure background.

Figure \ref{fig:psicontinj} shows the evolution of azimuthally
averaged poloidal flux lines in the $\{r,z\}$ plane. The initial
poloidal flux has a maximum of $\sim 0.37$, then the poloidal flux
increases to $\sim 0.9$ at $t=2.5$ when the injection finishes.
Magnetic fields continue to expand somewhat after the injection has
stopped.  Figure \ref{fig:enecontinj} shows the energy evolution for
different components.  The injected energy is essentially all in
magnetic fields, with a very small amount in the form of thermal and
kinetic energies of the injected plasmas associated with
$\dot{\rho}_{\rm inj}$. There is again a little numerical heating at
the late time (see the top two curves in Figure \ref{fig:enecontinj})
due to the existence of extremely low density regions.

Part of the injected magnetic energy is gradually converted into both
the kinetic energy and the changes in the thermal energy of the
surrounding plasmas.  The system is approaching a quasi steady state
with very slow or no changes in different energy components. We
interpret this as that the injected magnetic fields run out of their
"drive" since the magnetic field injection stops at $t=2.5$.
Subsequently, the magnetic fields have come to be in rough force
balance with the surroundings.  Of course the whole system can not be
steady, it is still expanding, though very slowly.  Figure
\ref{fig:finalcontinj} shows the distribution of magnetic fields,
density and gas pressure at $t=7$. A magnetically dominated cavity is
evident.

Figure \ref{fig:radcutcontinj} shows the radial distribution of
azimuthally averaged magnetic field components and pressures at $t=7$
and $z=3$. Again, we have a central column which is magnetically
dominated, with nearly the force-free condition. At a larger radius
($r \sim 1.5$), a small gas pressure gradient is present to balance
the residual magnetic pressure gradient. This hollow pressure profile
may be helpful to stabilize the whole structure.

It is interesting to note that more than half of the injected magnetic
energy still remains in the system for this set of parameters.
Compared with the impulsive injection case with $\alpha=20$, the two
systems have about the same initial ``free'' energy, and we can see
that the continuous injection case retains a much larger fraction of
the magnetic energy and gives a slightly larger magnetically dominated
volume.

\section{Discussions}
\label{sec:discu}

In this paper, we have presented a method of injecting free energy in
the purely magnetic form. The non-force-free nature of these magnetic
fields causes expansion in a self-pinched, collimated fashion. In
relating to astrophysical jets, we assume that the whole radio galaxy
can be divided into: a central region where the black hole accretion
system resides, a propagation region which contains the jets and
outflows supplied by the central region, and a termination region
where the lobes are formed. Based on this assumption, we suggest that
the approach taken here could be used to study both the propagation
and termination regions. We have especially concentrated on the
magnetically dominated regime, different from the previous studies
where energy is predominantly carried by the kinetic energy of the
flow.

The 3D MHD simulation results for a few relatively simple cases are
presented here, mostly for the purpose of illustrating the salient
features of the particular functional form of the injected magnetic
fields and how the subsequent evolution might be related to it.  As
such, we have only explored a limited set of parameters and their
influence on the magnetic field evolution. Many important issues are
not addressed here but will be presented in forthcoming papers. For
example, it will be important to know the stability of such magnetic
structures. The field line plots show that they are highly wounded,
suggesting kink unstable, yet the final distributions seem to suggest
that they remain axisymmetric and stable. The hollow pressure profile
might have a stabilizing influence, but more detailed studies are
required in order to draw firm conclusions on the stability. In
addition, how background environment can affect the propagation speed
and general morphologies, how the final configuration depends on the
injection rates, and what determines the total amount of magnetic
energy dissipation, etc.  It is thus premature to compare the current
simulation results directly with observations of radio jets/lobes,
though this is our ultimate goal.

\acknowledgments 

HL acknowledges useful discussions with D. Ryu and M. Nakamura.  This research
was performed under the auspices of the Department of Energy. It was
supported by the Laboratory Directed Research and Development Program
at LANL and by IGPP at LANL.

\end{document}